\documentclass[twocolumn,showpacs]{revtex4}

\usepackage{graphicx}
\usepackage{dcolumn}
\usepackage{bm}

\input epsf

\begin{document}

\title{\Large Modified Chaplygin Traversable Wormholes}

\author{\bf~Subenoy Chakraborty\footnote{schakraborty@math.jdvu.ac.in}
and~Tanwi~Bandyopadhyay}

\affiliation{Department of Mathematics,~Jadavpur
University,~Kolkata-32, India.}

\date{\today}

\begin{abstract}
The modified Chaplygin gas (MCG) is a strong candidate for the
unified model of dark matter and dark energy. The equation of
state of this modified model is valid from the radiation era to
the $\Lambda$CDM model. In early epoch (when $\rho$ was large),
dark matter had the dominant role while at later stages (when
$\rho$ is small), the MCG model behaves as dark energy. In this
work, we have found exact solution of static spherically
symmetric Einstein equations describing a wormhole for an
inhomogeneous distribution of modified Chaplygin gas. For
existence of wormhole solution, there are some restrictions
relating the parameters in the equation of state for MCG and the
throat radius of the wormhole. Physical properties and
characteristics of these modified Chaplygin wormholes are
analyzed in details.
\end{abstract}

\pacs{04.20.Gz,~~04.20.Jb,~~98.80.Es}

\maketitle

\section{\normalsize\bf{Introduction}}

Recent observational evidences of anisotropy in the Cosmic
Microwave Background radiation [1] and the data from type Ia
Supernovae SN 1997H redshift [2-4] suggest that the universe is
flat and is undergoing at present an accelerating phase preceeded
by a period of deceleration. These observational facts can not be
explained by the known ordinary (baryonic) matter or radiation.
There should be a significant amount of energy density of the
universe which should be an extraordinary non-baryonic matter
(dark matter) and energy (dark energy). Also, observational facts
indicate that, dark matter should be of the order of 25\% of the
critical density and dark energy contributes about $\frac{2}{3}$
of the the critical density [4-7]. Further, the universe in very
recent past (at redshift $z\lesssim1$), is dominated by uniformly
distributed dark energy with negative pressure and is probably
responsible for the present phase of acceleration of the
universe. In last few years, several models have been suggested
to incorporate the recent observational evidences and among them,
a single component perfect fluid having exotic equation of state,
known as Chaplygin gas [8-18] is of great interest. A further
generalized form is known as modified Chaplygin gas [the equation
of state is given in Section III, equation (12)]. The parameters
$A$, $B$ and $\alpha$ are universal positive constants. The
special choice $A=0$ and $\alpha=1$ corresponds to Chaplygin
gas,first introduced to describe lifting forces on a plane wing
in aerodynamics process [12]. Its generalization with the choice
$A=0$, $\alpha>0$ is known as generalized Chaplygin gas (GCG),
first introduced by Kamenshchik et al.[13] and Bento et al.[14].
For small energy density, the GCG model will have negative
pressure while at high energy density, the model behaves as an
almost pressureless fluid. However, MCG model, an extension of
GCG model [8,~15], can interpolate states between standard fluids
at high density and constant negative pressure at low energy. In
fact, the MCG model for $\gamma=\frac{1}{3}$ describes the
evolution from radiation epoch at early time to $\Lambda$CDM era
at late time (where the fluid has constant energy density and
behaves as a cosmological constant). This causes the present
accelerating phase of the universe [8,~9,~15].  Another point
that also goes against the GCG model is the measurement of the
velocity of sound. In GCG model, the velocity of sound is
negligible at early times and approaches the speed of light at
late times. But in cosmic evolution, sound velocity comparable to
the light velocity is not compatible.\\

From phenomenological view point, MCG model is interesting and
can be motivated by the brane world interpretation [14]. Further,
this model is consistent with various classes of cosmological
tests viz. gravitational lensing [17], gamma-ray bursts [18] as
well as the above mentioned observations. Moreover, the present
MCG model (also GCG model) is naturally constrained through
cosmological observables [16]. Also, for low energy density, the
present model has similarity with GCG model and so the equation
of state is that of a polytropic gas [15] with negative index.
Therefore, it is possible to have astrophysical implications of
the present model with an alternative way of restricting the
parameters [16].\\

As in MCG model, though the pressure is initially positive, but
with the evolution of the universe, the pressure becomes negative
and finally behaves as phantom energy. It is well known that
phantom energy violates null energy condition, a natural scenario
for the existence of traversable wormholes [19,~20]. This purely
theoretical ingredient possesses a peculiar property viz. "exotic
matter", whose stress-energy tensor violates the null energy
condition. This common fact of violation of null energy condition
motivates to study wormholes supported by MCG equation of state.
In this work, we shall construct static and spherically symmetric
traversable wormhole geometries, satisfying the MCG equation of
state. We shall consider matching of these wormhole geometries to
an exterior vacuum space time and examine physical properties and
characteristics of these solutions. We shall also study the
traversability conditions [19,~21] and measure the amount of
averaged null energy condition violating matter [22,~23] in
particular cases.\\

The remainder of our paper is organized as follows: Section II
provides basic equations for wormhole geometry. In Section III,
we have formulated the fundamental equations describing wormhole
in modified Chaplygin gas and the restrictions on the parameters
of the equation of state of MCG have been presented. Section IV
deals with asymptotic flatness of the wormhole geometry. The
non-flat (aymptotically) wormhole solutions are bounded by an
exterior vacuum space time for which, details of surface stresses
are shown. In section V, some specific wormhole solutions are
derived and their physical properties and characteristics are
explored. We have considered the choices viz. (i) a particular
relation between $\Phi$ and $b$ (in differential form), (ii)
constant redshift function, (iii) two typical choices of the
shape function and lastly (iv) isotropic pressure with a typical
form of $b(r)$. The traversability condition has been studied in
details for constant redshift function with a toy example. Also
for the linear form of $b(r)$, the amount of averaged null energy
condition violating matter has been estimated using "volume
integral quantifier". Finally, the paper ends with discussion and
concluding remarks in Section VI.\\

\section{\normalsize\bf{Basic Equations for Wormhole Geometry}}

A spherically symmetric and static wormhole is given in
Schwarzschild coordinates by the space time metric

\begin{equation}
ds^{2}=-exp~[2\Phi(r)]~dt^{2}+\frac{dr^{2}}{1-\frac{b(r)}{r}}
+r^{2}d\Omega^{2}
\end{equation}

where the two arbitrary functions of $r$ viz. $\Phi(r)$ and $b(r)$
are as usual termed as the redshift function and the shape
function respectively. The radial coordinate $r$ ranges over
$[r_{0},\infty)$ where the minimum value $r_{0}$ corresponds to
the radius of the throat of the wormhole. But there may be a
cut-off of the stress-energy tensor at any finite junction radius
$R$, where the interior space time matches to an exterior vacuum
solution.\\

The following properties need to be imposed for the existence of
a wormhole solution: [19,~20]

\begin{equation}
\text{I. As $r_{0}$ is the throat radius, so}~~ b(r_{0})=r_{0}
\end{equation}

\begin{eqnarray*}
\text{II. A flaring out condition of the throat, i.e}~~
\end{eqnarray*}
\begin{equation}
\frac{(b-b'r)}{b^{2}}>0,~\text{which at the throat simplifies
to}~~b'(r_{0})<1
\end{equation}

III.

\begin{equation}
b(r)<r~\text{for}~r>r_{0}
\end{equation}

~~~~~~~~~~(regularity of the metric coefficient)\\

IV. Traversability criteria: For traversable wormhole, there
should not be any horizons present i.e, $\Phi$ must be finite
everywhere.\\

Using the Einstein field equation $G_{\mu\nu}=8\pi T_{\mu\nu}$, in
an orthonormal reference frame ($G=C=1$), we obtain the following
set of equations

\begin{equation}
\rho(r)=\frac{b'}{8\pi r^{2}}
\end{equation}

\begin{equation}
p_{r}(r)=\frac{1}{8\pi}\left[\frac{2}{r}\left(1-\frac{b(r)}{r}\right)\Phi'
-\frac{b}{r^{3}}\right]
\end{equation}

and

\begin{eqnarray*}
p_{t}(r)=\frac{1}{8\pi}\left(1-\frac{b(r)}{r}\right)\left[\Phi''
+\Phi'\left(\Phi'+\frac{1}{r}\right)\right]
\end{eqnarray*}
\begin{equation}
-\frac{(b'r-b)}{2r^{2}} \left(\Phi'+\frac{1}{r}\right)
\end{equation}

in which $\rho(r)$, $p_{r}(r)$ and $p_{t}(r)$ are the energy
density, radial pressure and tangential pressure respectively.
The energy conservation equation gives

\begin{equation}
p_{r}'=\frac{2}{r}(p_{t}-p_{r})-(\rho+p_{r})\Phi'
\end{equation}

This equation can be interpreted as the hydrostatic equilibrium
equation for the material threading the wormhole.\\

The above Einstein equations can be rearranged to obtain metric
coefficients in terms of the components of the stress-energy
tensors as

\begin{equation}
b'=8\pi\rho(r)r^{2}
\end{equation}

\begin{equation}
\Phi'=\frac{b+8\pi
p_{r}r^{3}}{2r^{2}\left(1-\frac{b(r)}{r}\right)}
\end{equation}

The above flaring out condition implies that the wormhole should
be threaded with matter violating the null energy condition (NEC)
(a fundamental property of wormholes). Matter that violates the
NEC is termed as exotic matter [19,~24,~25]. The mathematical
form of the null energy condition is
$T_{\mu\nu}K^{\mu}K^{\nu}\geq0$ for any null vector $K^{\mu}$.
However, in an orthonormal reference frame, we may choose
$K^{\mu}=(1,1,0,0)$ so that the violation of NEC simplifies to
$\rho+p_{r}<0$ i.e,

\begin{equation}
\frac{1}{8\pi}\left[\frac{b'r-b}{r^{3}}+2\left(1-\frac{b}{r}\right)
\frac{\Phi'}{r}\right]<0
\end{equation}

Note that, if we evaluate at the throat considering the flaring
out condition and the finite character of $\Phi(r)$, the above
inequality (11) is automatically satisfied, i.e, the null energy
condition is violated.\\

\section{\normalsize\bf{Modified Chaplygin gas: The equation of state}}

The MCG is described by the equation of state [8,~9]

\begin{equation}
p_{r}=A\rho-\frac{B}{\rho^{\alpha}}~,~~~0<\alpha\leq1
\end{equation}

where the parameters $A$ and $B$ are two universal positive
constants. This equation of state shows a radiation era (when
$A=1/3$) at one extreme (when the matter density $\rho$ is very
large) and a $\Lambda$CDM model at the other extreme (when $\rho$
is infinitesimally small). However, at all stages it shows a
mixture. Further, somewhere in between the two extreme, there is
one stage when the pressure vanishes and the matter content is
equivalent to pure dust (this cosmological model can be
considered from the field theoretical point of view by
introducing a scalar field having self-interacting potential
[9]).\\

Now eliminating $\rho$ and $p_{r}$ among the equations (9), (10)
and (12), we have the relation between $\Phi$ and $b$ (in
differential form) as

\begin{equation}
\Phi'=\frac{\left[b+Arb'-\frac{rB(8\pi
r^{2})^{\alpha+1}}{(b')^{\alpha}}\right]}{2r^{2}\left(1-\frac{b}{r}\right)}
\end{equation}

Any solution of the metric (1) satisfying relation (13) is termed
as modified Chaplygin wormhole provided it obeys the restrictions
(mentioned in the previous section) for wormholes. In principle,
we now have four equations (8)-(10), (13) containing five
unknowns viz. two metric coefficients $\Phi(r)$ and $b(r)$ and
three physical quantities (components of stress-energy tensor)
$\rho$, $p_{r}$ and $p_{t}$. So we may restrict one of the above
unknown quantity (or a relation among them) for a specific
solution. In the present problem, we shall choose $b(r)$ or
$\Phi(r)$ so that the solution fulfills the criteria of
wormholes. Note that, as we shall consider the matter density
$\rho$ to be positive, so from equation (5), the shape function
$b(r)$ should be such that $b'(r)>0~\forall~r$.\\

Now at the throat $r=r_{0}$, the components of the stress-energy
tensor are (for the choice $\alpha=1$)

\begin{equation}
\left.
\begin{array}{ll}
\rho_{0}=\rho(r=r_{0})=\frac{1}{2A}\left[\sqrt{\left(\frac{1}{8\pi
{r_{0}}^{2}}\right)^{2}+4AB}-\frac{1}{8\pi {r_{0}}^{2}}\right]\\\\
p_{r_{(0)}}=-\frac{1}{8\pi {r_{0}}^{2}}\\\\
p_{t_{(0)}}=\frac{1-b'(r_{0})}{2r_{0}}\left(\Phi_{0}'+\frac{1}{r_{0}}\right)
\end{array}
\right\}
\end{equation}

The restriction $b'(r_{0})<1$ for the wormhole geometry restricts
the parameter $A$ and $B$ in MCG equation of state as

\begin{equation}
B<\frac{(1+A)}{(8\pi r_{0}^{2})^{2}}
\end{equation}

In the present context, the violation of null energy condition
i.e, $\rho+p_{r}<0$ demands the energy density to be restricted as

\begin{equation}
\rho^{\alpha+1}<\frac{B}{1+A}<\left(\frac{1}{8\pi
r_{0}^{2}}\right)^{2}
\end{equation}

This is not unusual in MCG cosmology models. Usually in FRW
cosmology, the evolution of $\rho$ is given by (see eqn. (8) in
Ref. [9])

\begin{equation}
\rho=\left[\frac{B}{1+A}+\frac{C}{a^{3(1+A)(1+\alpha)}}\right]^{(1/1+\alpha)}
\end{equation}

where $a$ is the scale factor in FRW model and $C$ is arbitrary
integration constant. So if $C$ is chosen to be positive, then
$\rho^{\alpha+1}>\frac{B}{1+A}$ i.e, dominant energy condition is
satisfied. But for the choice of negative values of $C$, the
restriction (16) is automatically satisfied, a fundamental
ingredient in wormhole physics. Moreover, the expression for the
velocity of sound is

\begin{equation}
v_{s}^{2}=\frac{\partial p_{r}}{\partial
\rho}=A+\frac{B\alpha}{\rho^{\alpha+1}}
\end{equation}

The condition that $v_{s}$ should not exceed the velocity of
light demands

\begin{equation}
\rho>\left(\frac{B\alpha}{1-A}\right)^{(1/1+\alpha)}
\end{equation}

Thus combining (16) and (19) we have the constraint

\begin{equation}
\frac{B\alpha}{1-A}<\rho^{\alpha+1}<\frac{B}{1+A}<\frac{1}{64\pi^{2}r_{0}^{2}}
\end{equation}

It should be mentioned that in the domain of exotic matter,
$\frac{\partial p}{\partial \rho}$ may not be interpreted as the
speed of sound due to the lack of knowledge of a microphysical
model of exotic matter. On the other hand, there are examples of
exotic behaviour such as Casimir effect and the false vacuum [26]
where $\frac{\partial p}{\partial \rho}<0$.\\

\section{\normalsize\bf{Asymptotic flatness and Stress-energy tensor cut-off}}

In the above, the basic structure of the solution has been shown
and the restrictions for the parameters involved are presented. A
solution for which $\Phi\rightarrow0$ and
$\frac{b}{r}\rightarrow0$ as $r\rightarrow\infty$, corresponds to
asymptotically flat space times. However, every solution may not
be related to asymptotically flat space time. Then the
stress-energy may be cut-off at a finite radial coordinate $R$
and the interior solution of metric (1) is matched to an exterior
vacuum space time at the junction interface $r=R$. The junction
$r=R$ may be a thin shell (if there are surface stresses at the
junction surface) or simply a boundary surface (zero surface
stresses). Hence for solutions corresponding to asymptotically
non-flat geometry, the MCG distribution is restricted to the
neighbourhood of the throat of the corresponding wormhole
solution.\\

If the exterior vacuum space time is chosen for simplicity as the
Schwarzschild space time i.e,

\begin{equation}
ds_{E}^{2}=-\left(1-\frac{2M}{r}\right)dt^{2}+\frac{dr^{2}}{(1-\frac{2M}{r})}
+r^{2}d{\Omega_{2}}^{2}
\end{equation}

then one must have\\

~~~~~~~~~~~~~~~~~~~$R>r_{b}=2M$  (event horizon).\\

Using the junction conditions due to Darmois-Israel formalism
[27], the components of the surface stresses of a dynamic thin
shell [28,~29] are

\begin{eqnarray*}
\sigma~\text{(surface energy density)}=-\frac{1}{4\pi
R}\left[\sqrt{1-\frac{2M}{R}+\dot{R}^{2}}\right.
\end{eqnarray*}
\begin{equation}
\left.-\sqrt{1 -\frac{b(R)}{R}+\dot{R}^{2}}\right]
\end{equation}

and

\begin{eqnarray*}
\mathcal{P}~\text{(tangential surface pressure)}=\frac{1}{8\pi
R}\left[\frac{1-\frac{M}{R}+\dot{R}^{2}+R\ddot{R}}{\sqrt{1
-\frac{2M}{R}+\dot{R}^{2}}}\right.
\end{eqnarray*}
\begin{equation}
\left.-\frac{(1+R\Phi')(1-\frac{b(R)}{R}
+\dot{R}^{2})+R\ddot{R}-\frac{\dot{R}^{2}(b-b'R)}{2(R-b)}}{\sqrt{1
-\frac{b(R)}{R}+\dot{R}^{2}}}\right]
\end{equation}

The above expressions for the components of stress-energy tensor
can be simplified to a great extent for the static case [30] by
imposing $\dot{a}=0=\ddot{a}$, where the overdot denotes
derivative with respect to the proper time $\tau$.\\

Further, if $M_{s}=4\pi r^{2}\sigma$ denotes the surface mass of
the thin shell (for static case), then the total mass $M_{T}$ of
the wormhole has the expression [21]

\begin{equation}
M_{T}=\frac{b(R)}{2}++M_{s}\left[\sqrt{1-\frac{b(R)}{r}}
-\frac{M_{s}}{2R}\right]
\end{equation}

Also for the static case, if we have the boundary surface i.e,
$\sigma=0=\mathcal{P}$, then one obtains the following
relationships

\begin{equation}
b(R)=2M~~~~\text{and}~~~~R=2M\left(\frac{\zeta-1/2}{\zeta-1}\right)
\end{equation}

As $R>2M$, so the redshift parameter $\zeta$ [$=1+R\Phi'(R)$] is
restricted by $\zeta>1$.\\

\section{\normalsize\bf{Specific wormhole solution}}

{\bf A.~~~~~~~\underline{A special relation between the redshift}}
{\bf \underline{function and the shape function:}}\\

Suppose we choose

\begin{equation}
\Phi'=\frac{A}{2}\frac{b'}{(r-b)}~~~\text{i.e,}~~\Phi=
\int\frac{A}{2}\frac{b'}{(r-b)}~dr
\end{equation}

Then from equation (13), the solution for $b$ becomes

\begin{equation}
b(r)=[r_{0}^{2}+\mu(r^{6}-r_{0}^{6})]^{1/2}~,~~~\mu=\frac{64\pi^{2}B}{3}
\end{equation}

with $3\mu r_{0}^{4}<1$ for $b'(r_{0})<1$.\\

Clearly, this solution does not correspond to an asymptotically
flat space time. So the MCG is confined in a neighbourhood of the
throat and the above solution can be matched to exterior vacuum
geometry (as described in Section IV). Further, the restriction
$b(r)<r$ yields a range of $r$ (for which wormhole solution is
possible) viz.

\begin{equation}
r_{0}<r<\left[\sqrt{\frac{4-3\mu
r_{0}^{4}}{4\mu}}-\frac{r_{0}^{2}}{2}\right]^{1/2}
\end{equation}

and is shown diagramatically in fig.(a).\\

\begin{figure}
\includegraphics[height=1.8in]{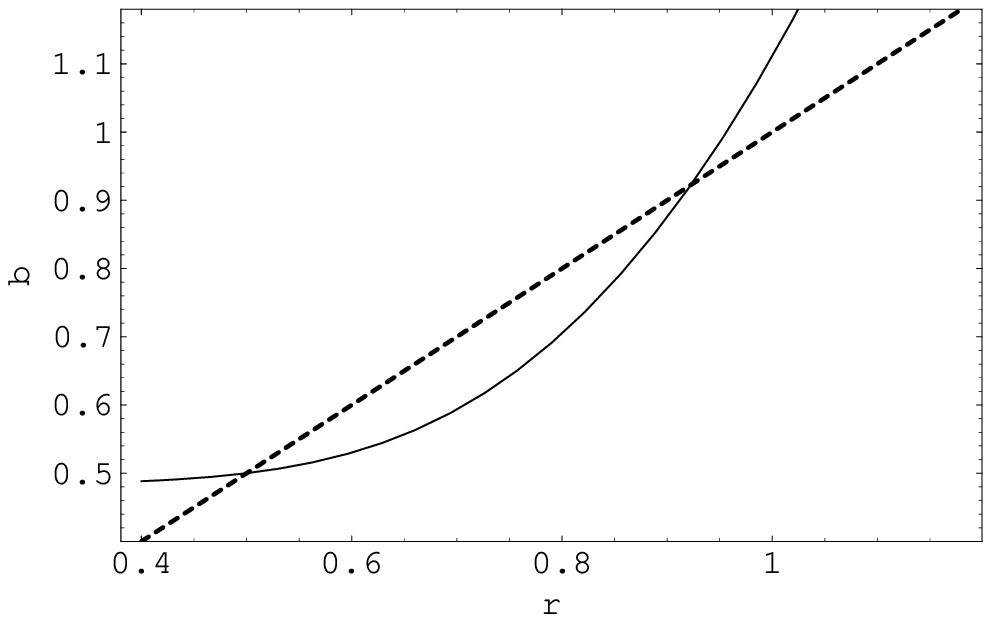}
\vspace{1mm}

\vspace{5mm} fig.(a) shows the variation of $b(r)$ (given by
eq.(27)) over the radial coordinate $r$ for $\mu=1$ and
$r_{0}=\frac{1}{2}$. \hspace{1cm} \vspace{6mm}
\end{figure}

{\bf B.~~~~~\underline{Constant redshift function:
~$\bm{\Phi'(r)=0$}}}\\

This choice of the redshift function gives (from (13)) the
differential equation in the shape function as

\begin{equation}
b(b')^{\alpha}+Ar(b')^{\alpha+1}-(8\pi)^{\alpha+1}Br^{2\alpha+3}=0
\end{equation}

One can immediately inspect that
$b(r)=r_{0}\left(\frac{r}{r_{0}}\right)^{3}$ is a particular
solution for $b(r)$ but it is not a wormhole solution as
$b'(r_{0})<1$ (flaring out condition) is violated. However, for
traversability criteria for constant $\Phi$, we need only $b'(r)$
at the throat, which can be obtained from (29) as (for $\alpha=1$)

\begin{equation}
b'(r_{0})=\frac{\sqrt{1+12A\mu r_{0}^{4}}-1}{2A}
\end{equation}

Note that for the flaring out condition we must have

\begin{equation}
3\mu r_{0}^{4}<1+A~,~~~~~\mu=\frac{64\pi^{2}B}{3}
\end{equation}

We shall now examine an important concept of wormhole physics
namely the traversability conditions for a human being to go
forward through the wormhole. The basic criteria for a smooth
journey is that the acceleration felt by the traveler should not
exceed earth's gravity $g_{e}$ (for details see ref.[19-21]) i.e,

\begin{equation}
\left|\left(1-\frac{b}{r}\right)^{1/2}e^{-\Phi}(\gamma
e^{\Phi})'\right|\leq g_{e}
\end{equation}

Moreover, for smooth journey across the wormhole, the tidal
acceleration also should not exceed the earth's gravitational
acceleration ($g_{e}$) and we have the following restrictions [19]

\begin{equation}
\left(1-\frac{b}{r}\right)\left[\Phi''+{\Phi'}^{2}-\frac{(b'r-b)}
{2r(r-b)}\Phi'\right]|{\eta^{\hat{1}}}'|\leq g_{e}
\end{equation}

\begin{equation}
\frac{\gamma^{2}}{2r^{2}}\left[v^{2}\left(b'-\frac{b}{r}\right)
+2(r-b)\Phi'\right]|{\eta^{\hat{2}}}'|\leq g_{e}
\end{equation}

Here the separation between two arbitrary parts of the traveler's
body is denoted by $|{\eta^{\hat{i}}}'|$ and is chosen to be $2m$
along any spatial direction in the traveler's
reference frame as in the literature [21,~25].\\

Suppose a human being travels with velocity $v$ along the radial
direction such that $v>0$ in $-l_{1}<l<l_{2}$ but $v=0$ at
$l=-l_{1}$ and $l_{2}$, where the two space stations are assumed
to be located (just outside the junction shell radius $R$) at
proper radial distances $l=-l_{1}$ and $l=l_{2}$ respectively.
($l=\int\frac{dr}{\sqrt{1-\frac{b}{r}}}$ is the proper radial
distance). Then the time of travel across the wormhole is given
by [19,~21,~31]

\begin{equation}
(\triangle T)_{t}={\int_{-l_{1}}}^{l_{2}}\frac{dl}{v\gamma}~~~
\text{and}~~~(\triangle T)_{0}={\int_{-l_{1}}}^{l_{2}}
\frac{dl}{ve^{\Phi}}
\end{equation}

where $\gamma=[1-v^{2}]^{-1/2}$ and $(\triangle T)_{t}$ is the
proper time interval of the journey recorded by the traveler's
clock, $(\triangle T)_{0}$ is the coordinate time interval
recorded by observers situated at the stations.\\

For simplicity, if we choose wormholes corresponding to
$\Phi=$constant (i.e, constant redshift), then for a traveler
moving with non-relativistic velocity (i.e, $\gamma\approx1$),
the restrictions (32) and (33) are identically satisfied while
the inequality (34), when evaluated at the throat, gives the
upper limit of the velocity as

\begin{equation}
v\leq r_{0}\sqrt{\frac{2g_{e}}{\left[1-b'(r_{0})\right]
|{\eta^{\hat{2}}}'|}}
\end{equation}

As a toy example if we choose
$b'(r_{0})=\frac{1}{2},~r_{0}=10^{2}$m, then one obtains
$v\approx4\times10^{2}$ m/s (choosing the equality sign) for the
maximum traversal velocity. Further, choosing the junction radius
$R\simeq10^{4}$m, the traversable time is $(\triangle
T)_{t}\simeq(\triangle T)_{0}\simeq\frac{2a}{v}\approx50$ sec.\\

{\bf C.~~~~\underline{Specific choice for the shape function:}}\\

~~~~~~~~~~~~~~~~~~~~~{\bf I.} $\bm{\underline{b(r)=r_{0}(\frac{r}{r_{0}})^{n}}}$\\

The particular choice of the shape function

\begin{equation}
b(r)=r_{0}\left(\frac{r}{r_{0}}\right)^{n}~,~~~~0<n<1
\end{equation}

satisfies the wormhole criteria viz.

\begin{eqnarray*}
(i)~b(r_{0})=r_{0}~,~~(ii)~b'(r)=n\left(\frac{r}{r_{0}}\right)^{n-1}~
\text{i.e,}
\end{eqnarray*}
\begin{equation}
b'(r_{0})=n<1~,~~(iii)~1-\frac{b(r)}{r}=1
-\left(\frac{r_{0}}{r}\right)^{1-n}<1
\end{equation}

Also
$\frac{b(r)}{r}=\left(\frac{r_{0}}{r}\right)^{1-n}\rightarrow0$
as $r\rightarrow\infty$.\\

Then solving equation (13) we have

\begin{eqnarray*}
\Phi(r)=\Phi_{0}+\frac{(1+An)}{2(1-n)}\left[\text{ln}\left|
\left(\frac{r}{r_{0}}\right)^{1-n}-1\right|-\text{ln}\left|
\frac{r}{r_{0}}\right|^{1-n}\right]
\end{eqnarray*}
\begin{eqnarray*}
-\frac{32\pi^{2}Br_{0}^{4}}
{n(1-n)}\left[\text{ln}\left|\left(\frac{r}{r_{0}}\right)^{1-n}
-1\right|+\alpha z\right.
\end{eqnarray*}
\begin{equation}
\left.+\frac{\alpha(\alpha-1)}{2}\frac{z^{2}}{2}+.........\right]
\end{equation}

where $z=\left(\frac{r}{r_{0}}\right)^{1-n}-1$,
$\alpha=\frac{5-n}{1-n}$ and $\Phi_{0}$ is the integration
constant. It can be seen easily that the above solution describes
a non traversable wormhole as there is an event horizon at the
throat $r=r_{0}$. However, it is possible to make the wormhole
solution traversable by imposing the restriction

\begin{equation}
\frac{1+An)}{2(1-n)}=\frac{32\pi^{2}Br_{0}^{4}} {n(1-n)}
\end{equation}

Then $\Phi(r)$ simplifies to

\begin{eqnarray*}
\Phi(r)=\Phi_{0}-\frac{1+An)}{2(1-n)}\left[\text{ln}\left|
\frac{r}{r_{0}}\right|^{1-n}+\alpha z\right.
\end{eqnarray*}
\begin{equation}
\left.+\frac{\alpha(\alpha-1)}{2}\frac{z^{2}}{2}+.........\right]
\end{equation}

As $\Phi(r)$ does not approach to a constant limit as
$r\rightarrow\infty$, so the solution is not asymptotically flat.
Hence we may match this interior solution to an exterior vacuum
space time at a junction radius $R$ and $\Phi_{0}$ can be
determined from the boundary conditions of $\Phi$ at the junction
interface.\\\\

~~~~~~~~~{\bf II.} $\bm{\underline{b(r)=r_{0}+d(r-r_{0})}}$\\\\

This typical choice of $b(r)$ satisfies as above all the criteria
to describe a wormhole provided $0<d<1$. It is clear that
$\frac{b}{r}\rightarrow0$ as $r\rightarrow\infty$ and the
solution for $\Phi$ becomes

\begin{eqnarray*}
\Phi(r)=\Phi_{0}+\frac{(1+Ad)}{2(1-d)}~\text{ln}|r-r_{0}|-\frac{1}{2}
~\text{ln}r-\frac{32\pi^{2}B}{d(1-d)}
\end{eqnarray*}
\begin{eqnarray*}
\left[r_{0}^{4}~\text{ln}|r-r_{0}|
+\frac{1}{4}~(r-r_{0})^{4}+4r_{0}^{3}(r-r_{0})\right.
\end{eqnarray*}
\begin{equation}
\left.+3r_{0}^{2}(r-r_{0})^{2}+\frac{4}{3}~
r_{0}(r-r_{0})^{3}\right]
\end{equation}

which clearly describes a non-traversable wormhole. However, the
choice\\

~~~~~~~~~~~~~~~~~~~~~$A=\frac{1}{d}
\left[64\pi^{2}Br_{0}^{4}-1\right]$\\

makes $\Phi$ finite at $r=r_{0}$ and the solution describes a
traversable wormhole (but not asymptotically flat). So the MCG is
confined around the throat up to a junction interface $r=R$.\\

We shall now examine how much of matter is in the space time,
which violates the averaged null energy condition. According to
Visser et al.[22,~23], this information can be obtained by "volume
integral quantifier" defined as

\begin{eqnarray*}
I_{v}=\int~[\rho(r)+p_{r}(r)]~dV
\end{eqnarray*}
\begin{equation}
={\int_{r_{0}}}^{R}(r-b)
\left[\text{ln}\left(\frac{e^{2\Phi}}{1-\frac{b}{r}}\right)\right]'dr
\end{equation}

Now using the solution (42) for $\Phi$ and the choice
$b(r)=r_{0}+d(r-r_{0})$, the above integral simplifies to

\begin{eqnarray*}
I_{v}=-(1+Ad)\left[\frac{a^{5}}{5}+a^{4}r_{0}+2a^{3}r_{0}^{2}
+2a^{2}r_{0}^{3}\right]
\end{eqnarray*}
\begin{equation}
+[1+Ad-r_{0}^{4}(1-d)]a
\end{equation}

where $a=R-r_{0}$ is the radial distance of the junction
interface from the throat. Thus as
$a\rightarrow0,~I_{v}\rightarrow0$. This shows that, similar to
the phantom [21] or GCG wormholes [25], it is possible to
construct a wormhole solution in MCG, where the amount of
modified Chaplygin gas is arbitrarily small. This conclusion is
in agreement with the results in the literature [21,~25]
regarding Chaplygin gas. The interesting point to be noted is
that, theoretically it is possible to construct wormholes
supported by infinitesimal amount of exotic fluids in cosmology,
which may responsible for the present accelerating phase of the
universe.\\

{\bf D.~~~~~\underline{Isotropic Pressure:}}\\

In case of isotropic pressure (i.e, $p_{r}=p_{t}$), the energy
conservation equation (8) simplifies to

\begin{equation}
\Phi'=-\frac{(A\rho^{\alpha+1}+B\alpha)}{\left[(A+1)\rho^{\alpha+1}
-B\right]\rho}~\rho'
\end{equation}

which on integration gives

\begin{eqnarray*}
\Phi=-\frac{A}{(A+1)(\alpha+1)}~\text{ln}\left|(A+1)\rho^{\alpha+1}-B\right|
\end{eqnarray*}
\begin{equation}
+\frac{\alpha}{\alpha+1}~\text{ln}\left|\frac{(A+1)\rho^{\alpha+1}}
{(A+1)\rho^{\alpha+1}-B}\right|+c
\end{equation}

The constant $c$ can be determined from the fact that, at the
throat $r=r_{0}$,\\

$\rho_{0}=\frac{1}{2A}\left[\sqrt{(\frac{1}{8\pi
r_{0}^{2}})^{2}+4AB} -\frac{1}{8\pi r_{0}^{2}}\right]$ [see
eqn.(14)] and $\phi(r=r_{0})=\phi_{0}$ (say), i.e, (for
$\alpha=1$)\\

$c=\phi_{0}+\frac{A}{2(1+A)}~\text{ln}|(A+1)\rho_{0}^{2}-B|
-\frac{1}{2}~\text{ln}\left|\frac{(1+A)\rho_{0}^{2}}{A+1)\rho_{0}^{2}-B}\right|$\\

Suppose we choose the shape function as [32]

\begin{equation}
b(r)=r_{0}+\gamma^{2}r_{0}\left[1-\frac{r_{0}}{r}\right]
\end{equation}

with $0<\gamma^{2}<1$.\\

Then from field equation (5) we have

\begin{equation}
\rho(r)=\frac{\rho_{1}}{r^{4}}~,~~~~~~~~~\rho_{1}=\frac{\gamma^{2}r_{0}^{2}}{8\pi}
\end{equation}

and the explicit expression for $\Phi$ becomes

\begin{eqnarray*}
\Phi(r)=c-\frac{A}{(A+1)(\alpha+1)}~\text{ln}\left|\frac{(A+1)
\rho_{1}^{\alpha+1}}{r^{4(\alpha+1)}}-B\right|
\end{eqnarray*}
\begin{equation}
+\frac{\alpha}{\alpha+1}
~\text{ln}\left|\frac{(A+1)\rho_{1}^{\alpha+1}}{(A+1)\rho_{1}^{\alpha+1}
-r^{4(\alpha+1)}B}\right|
\end{equation}

Clearly, the solution is not asymptotically flat and the radius
of the junction interface should be restricted by

\begin{equation}
R<\left[\frac{(A+1)\rho_{1}^{\alpha+1}}{B}\right]^{1/4(\alpha+1)}
\end{equation}

Finally, it should be mentioned that wormhole solution for
constant energy density is identical to that in ref. [25], so we
have not presented it here.\\

\section{\normalsize\bf{Discussion and Concluding remarks}}

The modified Chaplygin gas model is a unified model of dark
matter and dark energy, which causes the present accelerated
expansion of the universe. First of all in the context of
cosmology, the MCG model describes a smooth transition from a
decelerated expansion to the present epoch of cosmic acceleration
(i.e, a continuous transition from radiation era to $\Lambda$CDM
model). Also from the phenomenological view point, the model
describes unified macroscopic prescription of dark matter and
dark energy. Lastly, the model is favorable than the GCG model
(clearly presented in the introduction) and is consistent with
several cosmological tests.\\

In the present work, we have examined whether wormhole solutions
(traversable or non-traversable) are possible in the background
of MCG model. It is found that, for the existence of a generic
solution of a wormhole, the parameters in the equation of state
of MCG should be restricted as (for $\alpha=1$)\\

~~~~~~~~~~~~~~~~$\frac{B}{1+A}<(8\pi r_{0}^{2})^{2}$\\

and consequently, there is violation of null energy condition (a
fundamental requirement in wormhole physics). We have presented
four specific wormhole solutions and analyzed the physical
properties with characterization of these modified Chaplygin
wormholes. In the first solution, we have assumed a differential
relation between the redshift function $\Phi(r)$ and the shape
function $b(r)$. The wormhole solution is valid within a finite
range of $r$ which is matched with an exterior vacuum space time.
The range of $r$ for the wormhole solution is shown graphically.
For the choice of constant redshift function, no analytic form of
the shape function is possible. Only we have determined the
restriction in the parameters involved to satisfy the
traversability criteria. Then solutions are obtained for two
specific choices of the shape function. In both cases, the
wormhole solutions can be made traversable by imposing certain
restriction on the parameters involved. For the second choice, we
have calculated 'volume integral quantifier' ($I_{v}$) to have an
idea about the amount of matter violating averaged null energy
condition. We have seen that $I_{v}\rightarrow0$ as the junction
interface approaches to the throat. Lastly, we have obtained
wormhole solution for isotropic pressure. Here also due to
complicated form of $\Phi$ a function of energy density $\rho$,
the form of the shape function is assumed and solution is
confined in a finite restricted region. Finally, we have shown in
general the traversability condition for a human being. The
restrictions are simplified a great deal for constant redshift.
We have finally estimated the time of travel for a traveler
moving with non-relativistic speed. It should be mentioned that
the space stations should be located at large $r$, so that the
geometry at the stations is nearly flat [31]. As a concluding
remark, we may say that the MCG traversable wormholes have
far-reaching physical and cosmological implications. In addition,
for the use of interstellar shortcuts, an absurdly advanced
civilization may use them as time machines (probably violating
causality) [20,~33,~34] (for review see [35]).\\

{\bf Acknowledgement}:\\

The majority of the work has been done during a visit to IUCAA
under the associateship programme. The authors gratefully
acknowledge the warm hospitality and facilities of work at IUCAA.
Also T.B is thankful to CSIR, Govt. of India, for awarding Junior
Research Fellowship and to Anusua Baveja for fruitful discussions.\\

\end{document}